\documentclass[aip,rsi,preprint]{revtex4-1} 
\usepackage{graphicx}

\begin{document}

\title{Frequency doubling and stabilization of a Tm,Ho:YLF laser at 2051 nm to a high finesse optical cavity}


\author{Adam Kleczewski}%
  \email{adamklec@u.washington.edu}%
	\author{Matthew Hoffman}%
	\author{Eric Magnuson}%
	\author{Boris Blinov}%
	\author{Norval Fortson}%
	\affiliation{University of Washington, Department of Physics, 3910 15th Avenue NE Seattle, WA 98195, USA}%

\begin{abstract}
Light from a Tm,Ho:YLF laser operating at 2051 nm is frequency doubled then coupled into a high Fabry-Perot cavity with manufacturer quoted finesse ($\mathcal F$) in excess of 300,000.  The frequency of the laser is stabilized using the Pound-Drever-Hall (PDH) method.  A two channel feedback circuit allows for laser frequency stabilization with a bandwidth of 2 MHz.  This laser system has been used to drive the 6S$_{1/2}$ to 5D$_{3/2}$ transition in $^{138}$Ba$^+$, where a 2 ms laser-ion coherence times has been observed, demonstrating a linewidth of less than 500 Hz.  The intensity noise of this system is analyzed and found to be acceptable for several proposed experiments on a single trapped barium ion.
\end{abstract}

\maketitle

\section{Introduction}
Diode pumped solid state lasers with `eye safe' wavelengths near 2000 nm have received considerable attention for their potential to be used in LIDAR applications.  A number of groups have reported frequency stabilizing similar lasers by locking the laser frequency to spectral features in a molecular vapor cell \cite{laporta02, co2lock} as well as to a low-finesse ($\mathcal{F}\sim5$) Fabry-Perot etalon \cite{cavitylock}.  The long-term drifts in laser frequency observed in these systems were between tens of kilohertz and tens of megahertz.  While stabilizing the frequency of a $\sim$2000 nm laser to a broad molecular resonance may be sufficient for LIDAR applications, driving electric dipole-forbidden transitions requires a more robust stabilization scheme.  Here we report on the frequency stabilization of a diode pumped Tm,Ho:YLF laser to a high-finesse Fabry-Perot etalon.  The stabilized laser will be used to coherently drive the 6S$_{1/2}$ to 5D$_{3/2}$ electric quadrupole transition in Ba$^+$ as part of several proposed atomic physics experiments.

First, this laser will be used to selectively populate states in the metastable 5D$_{3/2}$ level of $^{137}$Ba$^+$.  Precision RF spectroscopy will be performed to measure the hyperfine splittings and fit for the magnetic octupole hyperfine coupling constant.  Because a barium ion has the electronic structure of an alkali atom, electronic wavefunctions for this system can be calculated with enough precision to extract a value for the nuclear magnetic octupole moment of $^{137}$Ba \cite{howell08}.  A similar measurement performed on the 6P$_{3/2}$ level of $^{133}$Cs implied a nuclear magnetic octupole moment 40 times larger than the predictions of the nuclear shell model \cite{tanner03}.  Second, the long lifetime of the 5D$_{3/2}$ level ($\tau \sim 80$s \cite{dehmelt97})and the fact that F=0 hyperfine sub-level does not couple to electric field gradients make this transition a possible candidate for use as an optical frequency standard \cite{sherman05}.  Finally, it is possible to drive the 6S$_{1/2}$ to 5D$_{3/2}$ transition in Ba$^+$ as a parity non-conserving electric dipole transition.  Fortson \cite{fortson93} has proposed an experiment to observe this effect in the light shift of the ground state energy levels of a single trapped barium ion caused by an off resonant 2051 nm laser.  

\section{Frequency Stabilization}
Robust frequency stabilization of a 2051 nm laser is a crucial component of all of the barium ion experiments described above.  The 2051 nm laser used for this work is a Tm,Ho:YLF diode pumped solid state laser manufactured by CLR Photonics (now a part of Lockheed Martin).  Its maximum output power was measured to be approximately 40 mW.  Changing the temperature of the laser resonator coarsely tunes the center wavelength of the laser.  Fine tuning of the laser frequency is achieved with a voltage sent to a piezo-electric actuator mounted on the output coupling mirror of the laser cavity.  The tuning piezo has a small-signal bandwidth of approximately 10 kHz.  

While reference cavities designed for use at visible and near-infrared wavelengths routinely have finesses greater than $10^5$, state-of-the-art high reflective coating technology for wavelengths near 2000 nm would limit the finesse of a 2051 nm cavity to approximately 20,000.  Additionally, commercial options for electro-optic modulators (EOMs), necessary for a PDH lock, and broadband acousto-optic modulators (AOMs), necessary for a wide tuning range, are not readily available. For these reasons we opted to frequency double our 2051 nm laser and reference the laser to a cavity designed for 1025 nm.

\subsection{Second Harmonic Generation of 2051 nm Light}
Implementing a PDH lock to a 1025 nm reference cavity requires that we generate at least 100 $\mu$W of 1025 nm light so that we can comfortably accommodate power losses from a double passed frequency shifting AOM.   Second harmonic generation (SHG) of light at a similar wavelength using a bulk periodically poled lithium niobate (PPLN) crystal in single pass configuration has been previously demonstrated with an efficiency of $10\%$ W$^{-1}$ \cite{galzerano03}.  The PPLN crystal used in this setup is 4 cm long and has nine parallel tracks with different poling periods.  We used a track with a poling period of 30.25 $\mu$m and maintained the crystal temperature at $108\,^{\circ}\mathrm{C}$ for optimum phase matching.  However, in single pass configuration we were only able to reach a maximum conversion efficiency of $1.25\%$ W$^{-1}$.  The reason for this lower efficiency is not clear.

1025 nm SHG efficiency is increased substantially by placing the temperature controlled PPLN crystal inside a ``bow-tie" enhancement cavity (Fig.~\ref{fig:schematic}).  The enhancement cavity consists of two flat mirrors and two curved mirrors each with a radius of curvature of 250 mm arranged in a bow-tie configuration.  The curved mirrors are separated by 325 mm and the PPLN crystal is placed at the midpoint where the beam comes to a 150 $\mu$m waist.  These relatively long mirror separations were necessary to ensure that the beam was not clipped by the 0.5 mm square PPLN channels.  This crystal is AR coated for 2051 nm, as well as 1025 nm, and has been polished to a 1 degree wedge to prevent the crystal from acting as an intra-cavity etalon.  The length of the enhancement cavity is locked to the wavelength of the 2051 nm laser by sending an error signal derived using the Hansch-Couillaud method\cite{hansch80} to a piezo-electric actuator mounted on one of the flat cavity mirrors.  Our coupling efficiency into the enhancement cavity is approximately 75\%.  We estimate the cavity build up of 2051 nm light to be approximately 20.  Sending the full 40 mW output of the Tm,Ho:YLF laser to the enhancement cavity we were able to generate 2.5 mW of 1025 nm light.  When the 55 MHz AOM is included in the 2051 nm beam path, approximately 20 mW is sent to the enhancement cavity and we are able to generate approximately 500 $\mu$W of 1025 nm light.

\subsection{PDH Stabilization with Two-Channel Feedback}
After the enhancement cavity, the 1025 nm beam makes a double pass through a frequency shifting AOM with a center frequency of 200 MHz.  This AOM allows for a continuous tuning range of approximately 80 MHz at the 2051 nm wavelength.  The beam then passes through a resonant EOM to generate PDH sidebands at 20 MHz \cite{hall83} and then through a series of mode matching lenses before coupling into the reference cavity.  

Our reference cavity consists of two mirrors held 77.5 mm apart by a spacer made from ultra-low expansion (ULE) glass of a design developed at JILA \cite{notcutt05}.  The mirrors are coated for high reflectivity at the 1025 nm wavelength.  The input mirror is flat and the output mirror has a radius of curvature of 500 mm.   The cavity is mounted vertically inside a vacuum chamber and maintained at a pressure of $10^{-8}$ Torr with an ion pump.  The vacuum chamber is enclosed inside a temperature stabilized box to further reduce cavity length variations.  Cavity ring down measurements indicate the cavity finesse to be greater than 300,000.

Light rejected from the cavity is separated from the incident beam using a quarter waveplate and a polarizing beam splitter.  The intensity of the rejected beam is detected with a high-bandwidth amplified photodetector (PD).  A PDH error signal is derived from the photodetector output using a double balanced mixer.

 \begin{figure}[t]
	\centering
		\includegraphics{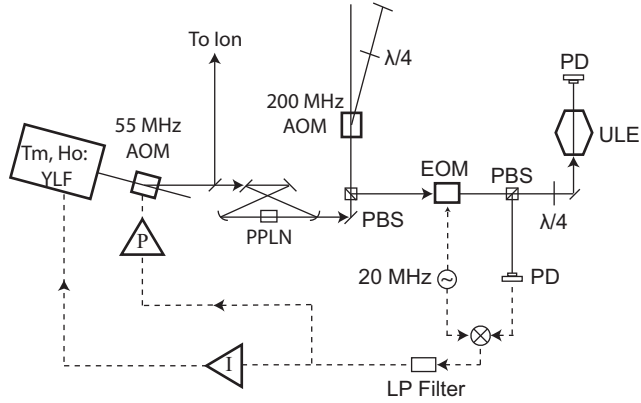}
		\caption{Light from a 2051 nm Tm,Ho:YLF laser passes through an AOM and is  frequency doubled with a PPLN crystal in a bow-tie cavity.  The 1025 nm light is then frequency shifted by a double-passed AOM, and stabilized to a high finesse reference cavity using the PDH method \cite{hall83}.  An EOM driven by a 20 MHz source is used to excited the sidebands necessary for the PDH lock. After passing through a low-pass (LP) filter, high bandwidth proportional feedback (P) is sent to the first AOM, while low bandwidth integral feedback (I) is amplified and sent to the tuning piezo inside the laser head.}
		\label{fig:schematic}
\end{figure}

Frequency stabilization of the Tm,Ho:YLF laser frequency is achieved by splitting the PDH error signal into a high bandwidth and a low bandwidth channel.  The low bandwidth channel consists of an analog integrator circuit (I in Fig.~\ref{fig:schematic}) followed by a high-voltage amplification stage.  The amplified signal is sent to the tuning piezo inside the laser head.  The bandwidth of this channel is approximately 10 kHz.  Higher bandwidth feedback is necessary to maintain a robust lock to the $\sim$7 kHz wide TEM00 mode of the ULE cavity.  To achieve this, the PDH error signal in the high bandwidth channel is sent through an analog amplification stage that provides variable proportional gain then to the control input of the voltage controlled oscillator (VCO) used to drive the 55 MHz AOM at the output of the laser.  Both the VCO and the AOM have 2 MHz modulation bandwidths.  With the gains of both feedback channels set correctly are we able to maintain a robust lock to the ULE cavity.  With the 2051 nm laser locked to the ULE cavity we were recently able to observe 2 ms laser-ion coherence times on the 6S$_{1/2}$ to 5D$_{3/2}$ transition in $^{138}$Ba \cite{klec11}, indicating a laser linewidth of less than 500 Hz.

\section{Intensity Noise}
\begin{figure}[t]
	\centering
		\includegraphics{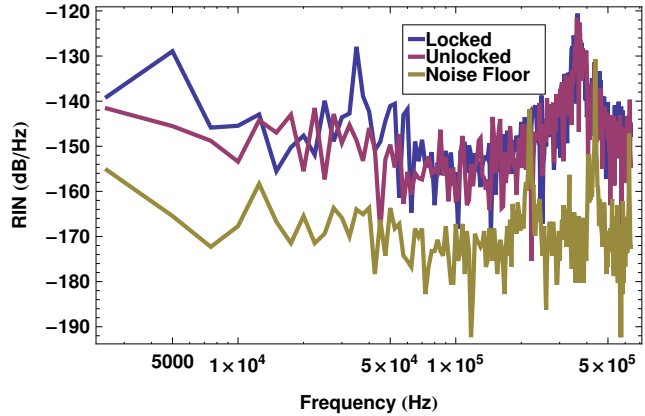}
		\caption{Relative intensity noise (RIN) spectrum on the Tm,Ho:YLF laser while free running and while locked to the ULE reference cavity.}
	\label{fig:intensitynoise1}
\end{figure}

One side effect of the frequency stabilization scheme presented here is that frequency noise canceled in the high bandwidth channel is converted into a small amount of intensity noise after the 55 MHz AOM.  The output amplitude of the VCO used to drive the AOM varies slightly with frequency, as does the efficiency of the AOM itself.  When the fast feedback circuit corrects for a fluctuation in laser frequency by changing the frequency of the VCO, the optical power after the AOM changes.  We measured the intensity noise on the 2051 nm beam that was sent to the barium ion using a high bandwidth photodetector, then displayed the frequency spectrum using a spectrum analyzer.  Relative intensity noise measurements of the free running laser and stabilized laser are shown in Figure \ref{fig:intensitynoise1}.  The peak at 400 kHz corresponds to the oscillation relaxation frequency of Tm,Ho:YLF, which is discussed in \cite{laporta02}.  The peak in the locked signal is at $\sim 35$ kHz is introduced by the frequency lock.  The total relative intensity noise in the 1 to 600 kHz frequency band increases from -104.7 to -100.2 dB when we lock the laser to the ULE cavity.  This excess intensity noise as well as the intensity noise at the relaxation frequency could be canceled using the intensity noise cancellation\cite{laporta02}, however we do not expect intensity noise to pose a serious problem for the planned experiments.

\section{Conclusions}
Efficient SHG of 1025 nm light was demonstrated using a PPLN crystal inside a ``bow-tie"  enhancement cavity.  Using two channel feedback circuit, the frequency of a 2051 nm Tm,Ho:YLF laser was locked to a high finesse reference cavity using the Pound-Drever-Hall technique, resulting in a sub-kilohertz laser linewidth.\\

\section*{Acknowledgments}
The authors wish to thank Jeff Sherman and Yonatan Cohen for early work on this project.  This research was supported by National Science Foundation grant PHY-09-06494.

\end{document}